\documentclass[3p,times,procedia]{elsarticle}
\usepackage{nupha_ecrc}
\usepackage{upgreek}

\volume{00}

\firstpage{1}

\journalname{Nuclear Physics A}

\newcommand{ \be }{\begin{equation}}      
\newcommand{ \ee }{\end{equatiton}}      
\newcommand{ \bea }{\begin{eqnarray}}      
\newcommand{ \eea }{\end{eqnarray}}

\usepackage{color}

\newcommand{ \sNN }{\sqrt{s_{NN}} }           

\newcommand{\pT}{$p_{T}$ }
\newcommand{\pp}{$p+p$ }

\newcommand{\AuAu}{Au+Au }
\newcommand{\raa}{$R_{AA}$ }
\newcommand{\ncoll}{$N_{coll}$ }

\newcommand{\nbin}{$N_{bin}$}

\runauth{}

\jid{nupha}

\jnltitlelogo{Nuclear Physics A}

\usepackage{amssymb}
\usepackage[figuresright]{rotating}

\begin{document}

\begin{frontmatter}



\dochead{}
%% e.g. \dochead{17th International Conference on Dynamical Processes in Excited States of Solids}

\title{Nuclear Modification Factor of $D^0$ Meson in Au+Au Collisions at $\sNN$ = 200 GeV}


\author{Guannan Xie (for the STAR\fnref{col1}  Collaboration)}
\fntext[col1] {A list of members of the STAR Collaboration and acknowledgments can be found at the end of this issue.}

\address{
Lawrence Berkeley National Laboratory, Berkeley, CA 94720, USA

University of Science and Technology of China, Hefei, 230026, China
}

\begin{abstract}
Heavy-flavor quarks are dominantly produced in initial hard scattering processes and experience the whole evolution of the system in heavy-ion collisions at RHIC energies. Thus they are suggested to be an excellent probe to the medium properties through their interaction with the medium. In this proceedings, we report our first measurement of $D^0$ production via topological reconstruction using STAR's recently installed Heavy Flavor Tracker (HFT). We also report our new measurement of Nuclear Modification Factor ($R_{AA}$) of $D^0$ mesons in central Au+Au collisions at $\sNN$ = 200 GeV as a function of transverse momentum ($p_{T}$). New results confirm the strong suppression at high \pT with a much improved precision, and show that the \raa at high \pT are comparable with light hadrons ($\pi$) and with D meson measurements at the LHC. Furthermore, several theoretical calculations are compared to our data, and with charm diffusion coefficient 2${\pi}TD_{S}$ $\sim$ 2-12 can reproduce both the $D^0$ \raa and $v_2$ data in \AuAu collisions at RHIC.
\end{abstract}


\end{frontmatter}

% \linenumbers

\section{Introduction}
The mass of charm quark is significantly larger than those of light quarks, $\Lambda_{QCD}$, and the QGP temperature at RHIC energies ($m_{c} >>  m_{u,d,s}, \Lambda_{QCD}, T_{QGP(RHIC)}$). Therefore charm quarks are dominantly produced in the early stage of the collision in hard scattering processes at RHIC. They experience the whole evolution of the system and offer unique information for the study of hot and dense strongly-coupled Quark-Gluon Plasma (sQGP) matter.

Charm production has been systematically measured in \pp($\overline{p}$) collisions in various experiments. Due to the large quark mass, charm production in \pp collisions is expected to be calculable with a good precision in perturbative QCD. Figure 1 (left) shows the charm differential cross-section at midrapidity versus transverse momentum in \pp collisions at $\sqrt{s}= 200$ GeV-7 TeV~\cite{star1, star2, cdf, alice}. Experimental data are compared with Fixed-Order Next-to-Leading-Log (FONLL) pQCD calculations shown as grey bands~\cite{fonll}. Within uncertainties, FONLL pQCD calculations describe the data over a broad range of collision energies. The precision of the experimental data allows to constrain the theoretical uncertainty in the pQCD calculations.

At RHIC energies, charm quarks are produced mostly via initial hard scatterings. This has been proved by charm total cross sections measured from different collision systems. Figure 1 right panel shows that, charm total cross section follows a number-of-binary-collision (\nbin) scaling~\cite{star1, fonll, star3, star4, nlo}. 

The modification of the charmed meson production is quantified with a Nuclear modification factor $R_{AA}$. \raa is calculated as the ratio between the $D^0$ invariant yield in \AuAu collisions to the \pp data scaled by \nbin. Our previous result shows that the $D^0$ \raa as a function of \pT is significantly different from unity~\cite{star4}. 

\begin{figure}[htbp]
\hspace{+1.2cm}
\begin{minipage}[b]{0.35\linewidth}%0.5
\begin{center}
\includegraphics[width=\textwidth]{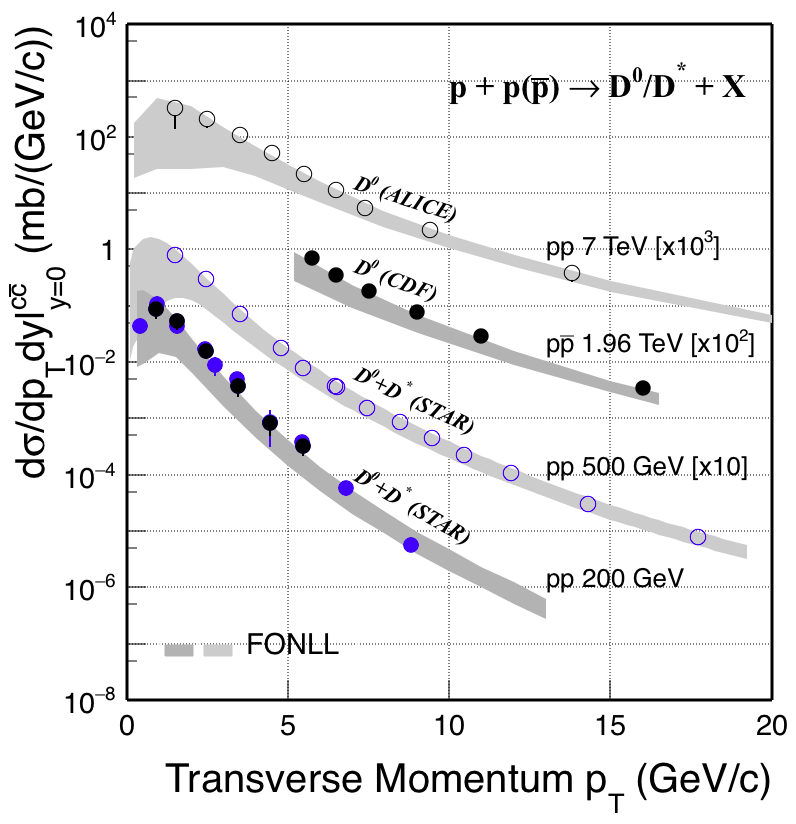}
\end{center}
\end{minipage}
\hspace{+1.2cm}
\begin{minipage}[b]{0.4\linewidth}%0.57
\begin{center}
\includegraphics[width=\textwidth]{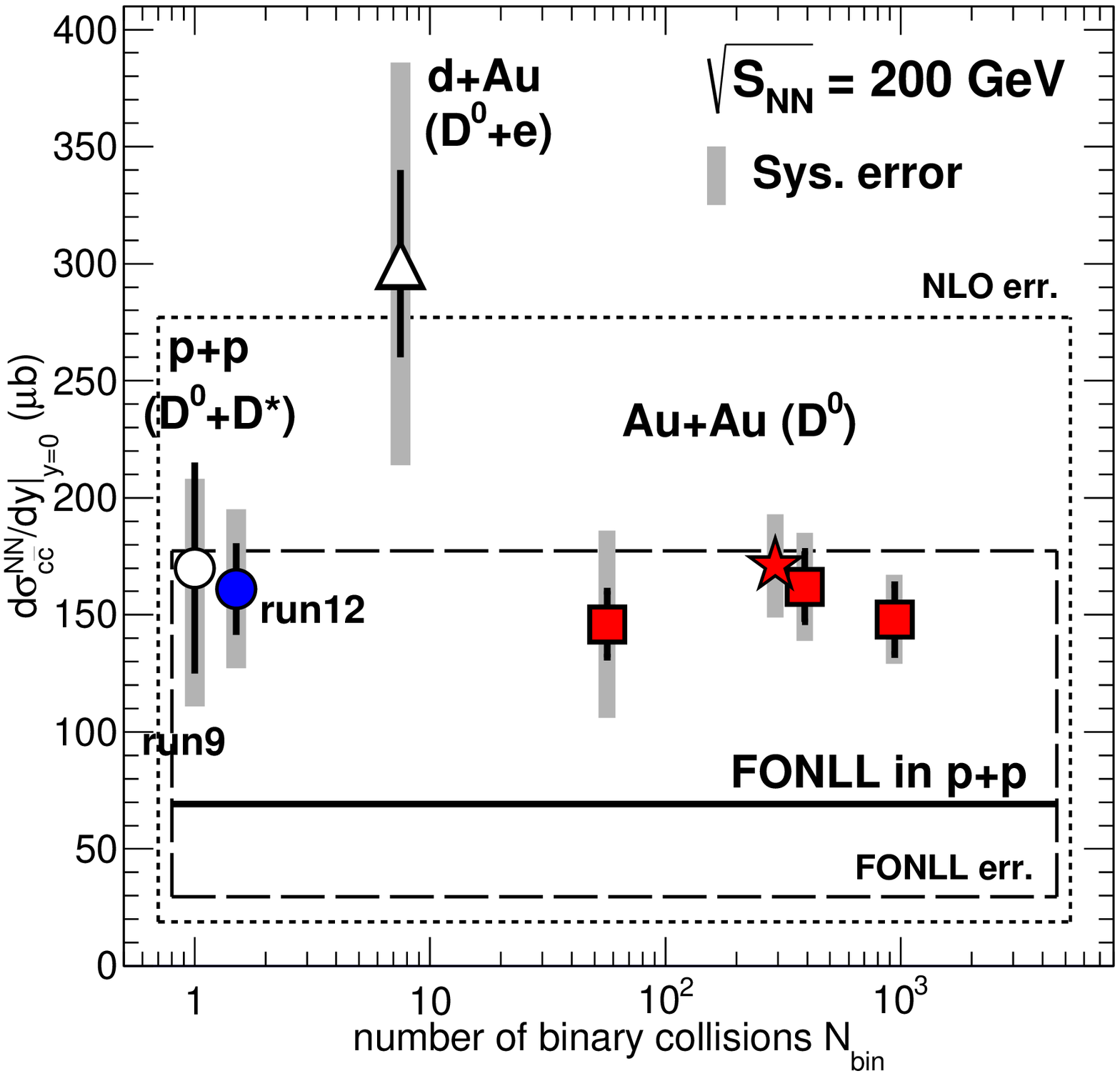}
\end{center}
\end{minipage}
\put(-280, 30) {\footnotesize \color{blue} STAR Preliminary}
\put(-140, 30) {\footnotesize \color{blue} STAR Preliminary}
\caption{ (Left) Charm quark pair production cross section vs. \pT at mid-rapidity in \pp($\overline{p}$) collisions at $\sqrt{s}= 200$ GeV-7 TeV. FONLL pQCD calculations are shown as shaded bands~\cite{fonll}. (Right) Charm cross section at mid-rapidity from \pp to central \AuAu collisions from STAR.} 
\label{fig:cc_pp_xSection}
\end{figure}

\section{Experiment and Analysis}

The STAR experiment is a large-acceptance multi-purpose detector which covers full azimuth and pseudorapidity of $|\eta| < 1$. In this analysis, the data were taken by the STAR experiment using the newly installed Heavy Flavor Tracker (HFT) in the year 2014 RHIC run. The HFT is a high resolution silicon detector, which consists of three subsystems: two conventional strip detectors and one pixel detector using the state-of-the-art Monolithic Active Pixel Sensors (MAPS) technology~\cite{star5}. The HFT is designed to greatly improve open heavy flavor hadron measurements by the topological reconstruction of secondary decay vertices. 

The data sample used in this analysis is $\sim$780M minimum bias events taken in Au+Au collisions at $\sNN$ = 200 GeV with the HFT. Events are required to have reconstructed collision position within 6 cm along the beam direction from the detector center. $D^0$ and $\overline{D^0}$ are reconstructed in the hadronic $K^{\mp}\pi^{\pm}$ channel, with a branching ratio of $\sim3.9\%$ and a lifetime of $c\tau\sim123$ $\upmu$m. Kaons and pions are identified via a combination of the energy loss dE/dx measured by the Time Projection Chamber and $\beta$ measured by the Time-Of-Flight detector~\cite{stardetector}. Secondary vertices are reconstructed as the middle point at the Distance of the Closest Approach (DCA) between the trajectories of two daughter particles. With the HFT, several topological cuts are used to greatly reduce the combinational background. Rectangular topological cuts are optimized in each $D^0$ \pT bins using the Toolkit for Multivariate Data Analysis (TMVA) package for best significance of $D^0$ signal.

Figure 2 shows the $D^0$ invariant mass distributions in two \pT bins. Compared with our former published result, the $D^0$ signal significance scaled to the same number of events has been improved by a factor of 4. The remaining combinatorial background is estimated with like-sign and mixed-event methods. The TPC efficiency is calculated by embedding simulated tracks into real event background. The HFT related and topology cut efficiencies are from $D^0$ decay simulation based on the HFT matching to TPC ratio and track pointing resolution directly from data. The results for \pT $<$ 2 GeV/c are being finalized, and we show the results below for $D^0$ \pT $>$ 2 GeV/c.

\begin{figure}[htbp]
\hspace{+1.2cm}
\begin{minipage}[t]{0.335\linewidth}
\begin{center}
\includegraphics[width=\textwidth]{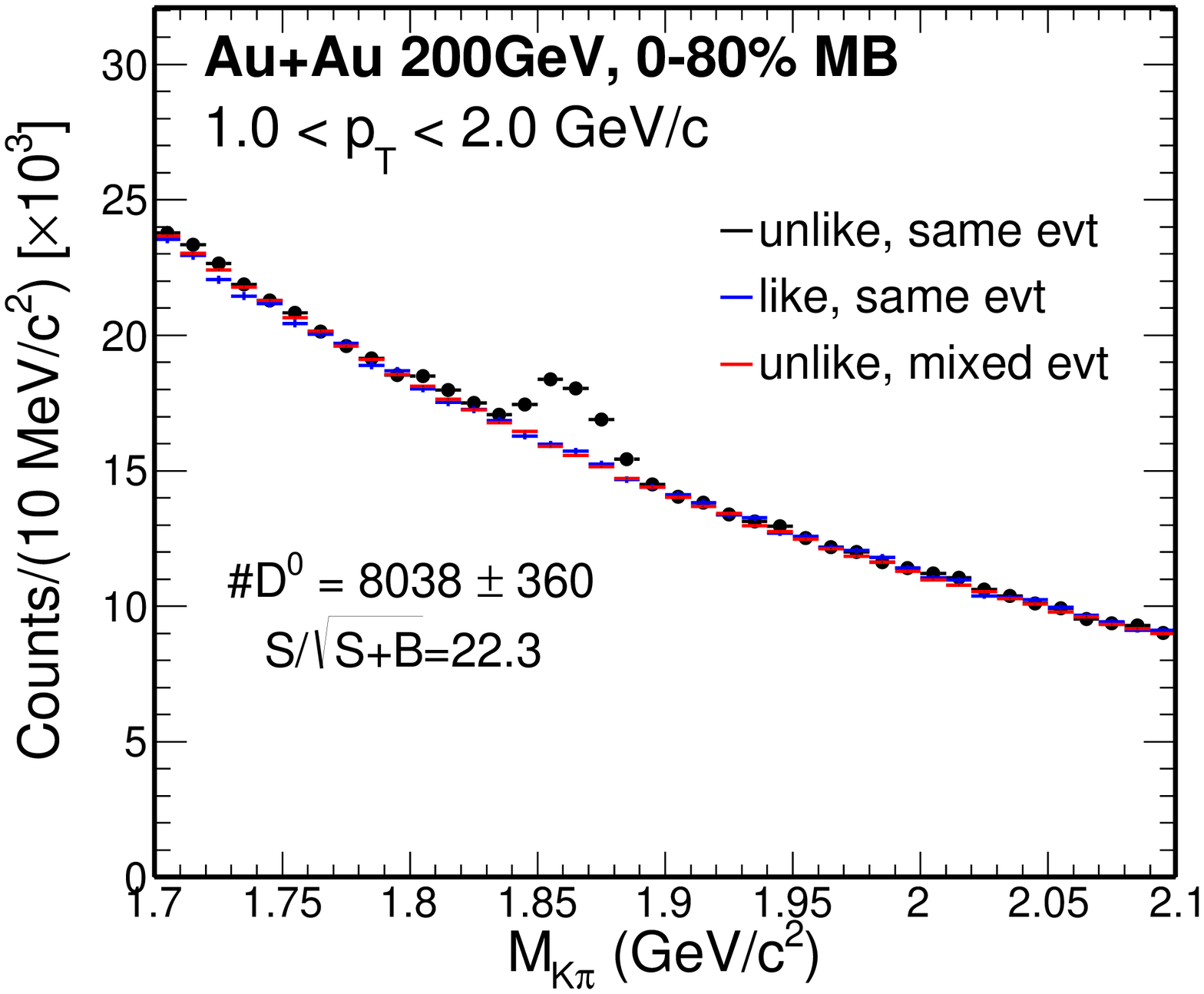}
\end{center}
\end{minipage}
\hspace{+1.2cm}
\begin{minipage}[t]{0.335\linewidth}
\begin{center}
\includegraphics[width=\textwidth]{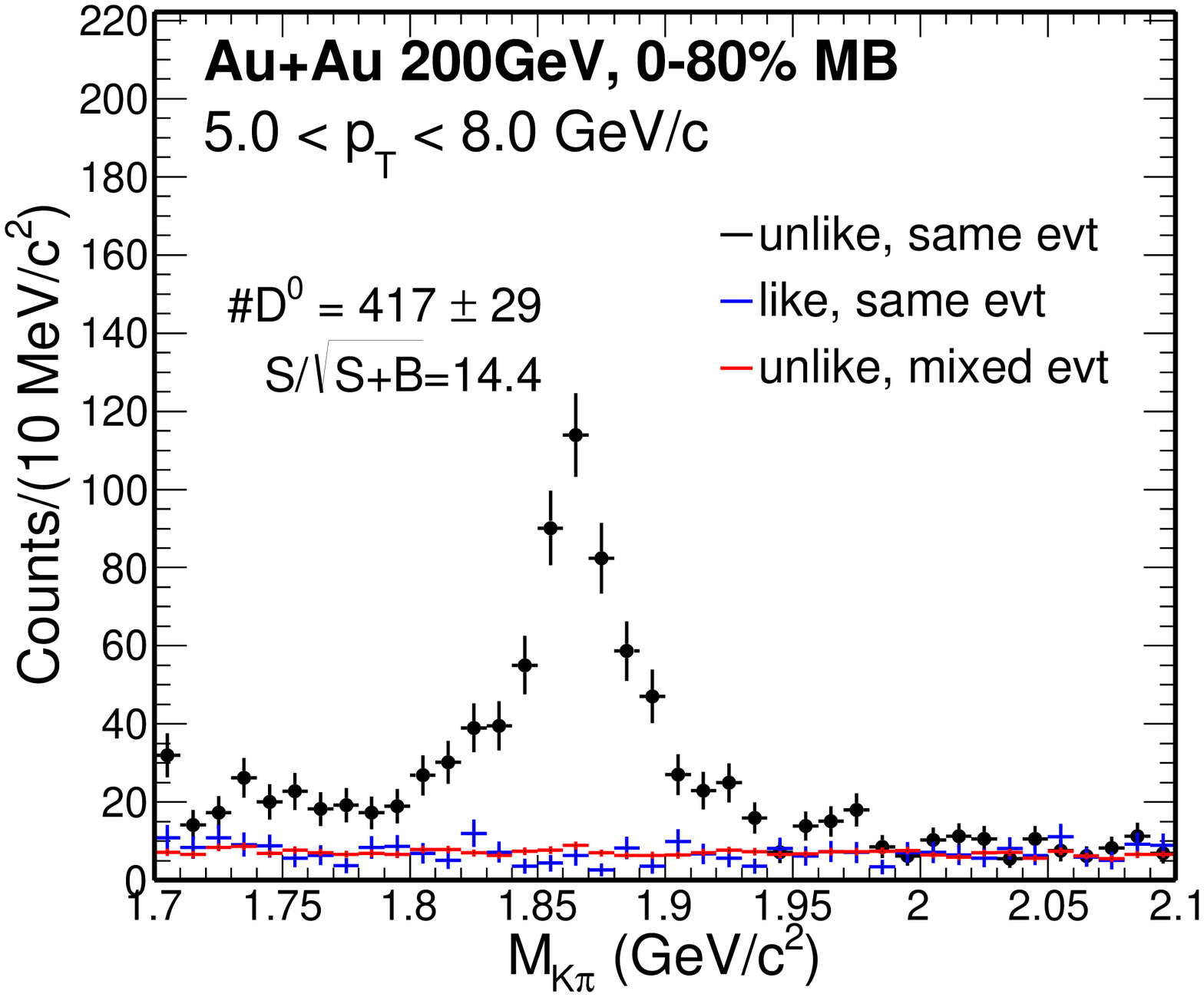}
\end{center}
\end{minipage}
\put(-305, 28) {\footnotesize \color{blue} STAR Preliminary}
\put(-65, 28) {\footnotesize \color{blue} STAR Preliminary}
\caption{Invariant mass spectra of $K\pi$ pairs for $1 < p_{T} < 2$ GeV/c (Left) and $5 < p_{T} < 8$ GeV/c (Right). The black points are unlike-sign pairs with the $D^0$ signal. The blue and red points show the like-sign and mixed-event background respectively.}
\label{fig:D0Signal}
\end{figure}

\section{Physics Results and Discussion}
Although the total charm cross section follows $N_{bin} $ scaling, the \pT spectrum is significantly modified in \AuAu collisions~\cite{star4}. $D^0$ invariant yield (left) and nuclear modification factor \raa(right) in the most central \AuAu collisions at $\sNN$ = 200 GeV are shown in Figure 3. The new $D^0$ invariant yields from the HFT (blue) are consistent with published data (red) with significantly improved precision~\cite{star4}.

\begin{figure}[htbp]
\hspace{+1.2cm}
\begin{minipage}[t]{0.335\linewidth}
\begin{center}
\includegraphics[width=\textwidth]{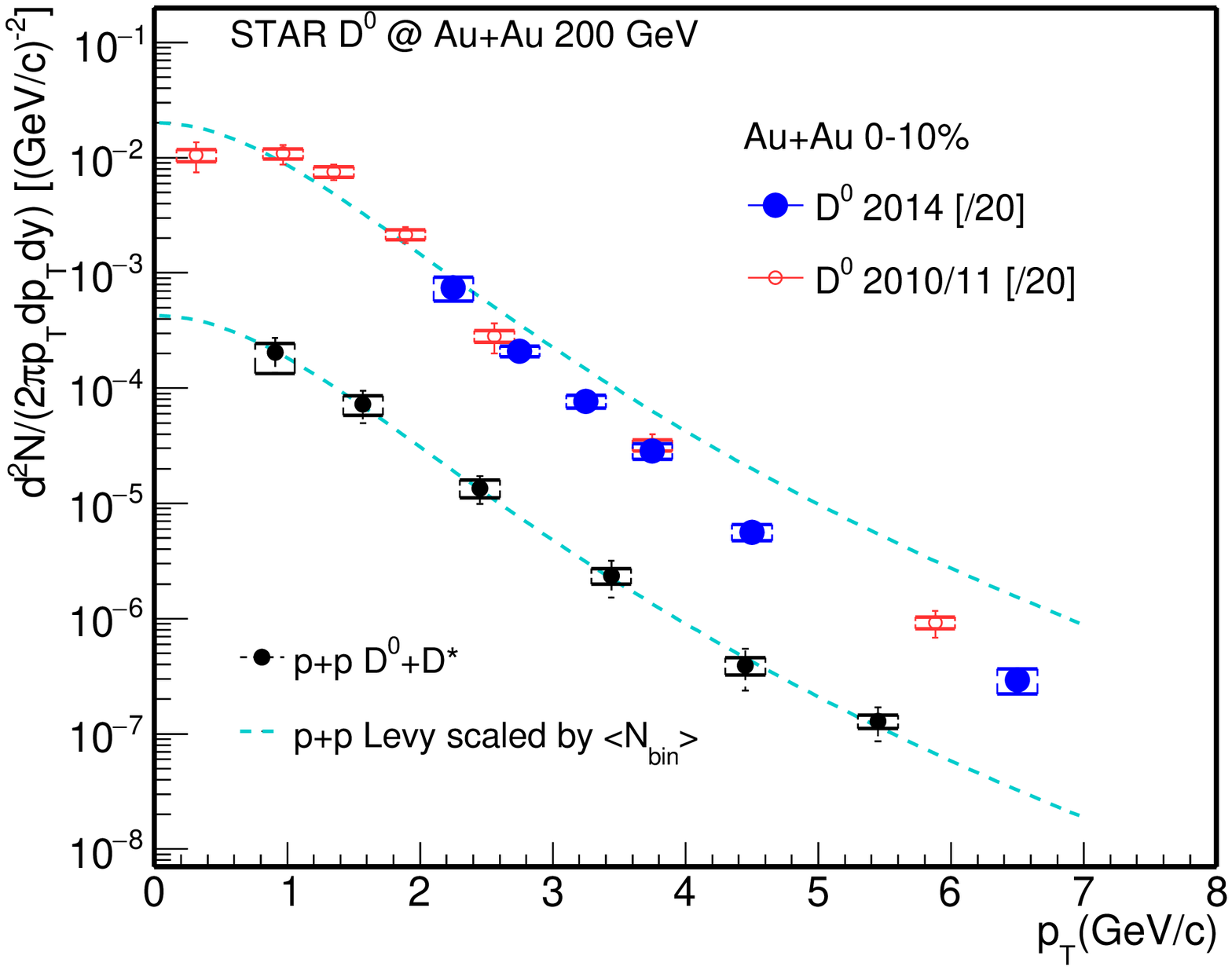}
\end{center}
\end{minipage}
\hspace{+1.2cm}
\begin{minipage}[t]{0.335\linewidth}
\begin{center}
\includegraphics[width=\textwidth]{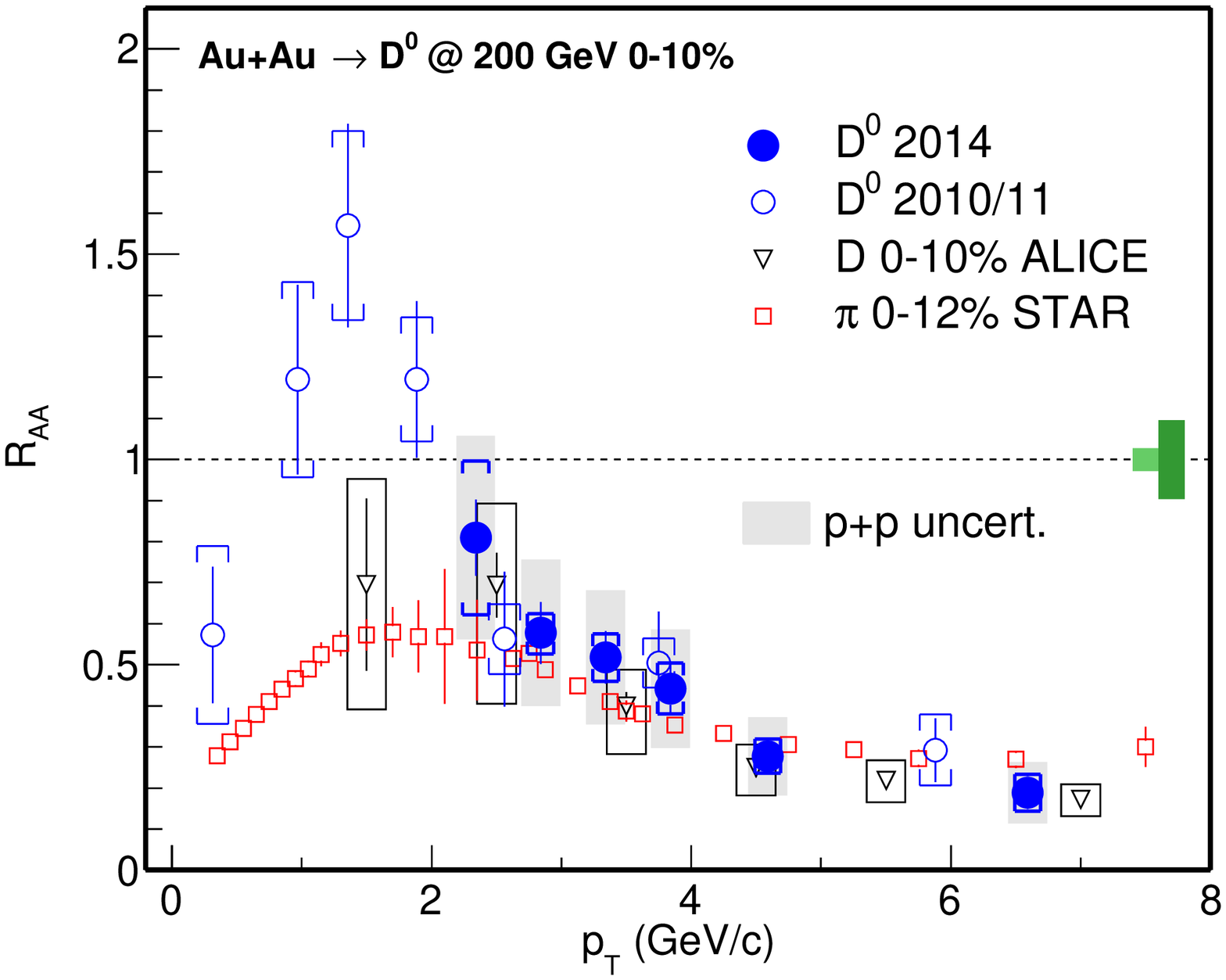}
\end{center}
\end{minipage}
\put(-303, 18) {\footnotesize \color{blue} STAR Preliminary}
\put(-122, 18) {\footnotesize \color{blue} STAR Preliminary}
\caption{$D^0$ meson invariant yields (Left) and nuclear modification factor \raa(Right) in most central Au+Au collisions at $\sNN$ = 200 GeV. Also shown in the left panel are $D^0$ yields in \pp collisions at $\sNN$ = 200 GeV fitted with a Levy function. In the right panel, vertical bars on data points indicate statistical uncertainties, while the brackets are systematic uncertainties in \AuAu collisions. The grey bands are uncertainties from \pp baseline. \raa of D mesons from LHC and $\pi$ from RHIC are also shown. There are two vertical boxes around unity from left to right related to \AuAu \ncoll and \pp normalization uncertainties.}
\label{fig:D0Spectra}
\end{figure}

Figure 3 (right panel) shows the \raa results from the most central (0-10\%) \AuAu collisions. The new results from the HFT are consistent with the published ones in the measured \pT range. Furthermore, they have highly improved precision in \AuAu collisions. The grey bands show uncertainties from the \pp baseline from our previous measurement before the HFT installation~\cite{star1}, which is expected to be improved as well with the HFT data taken in the year 2015 RHIC run. The \raa shows a strong suppression at high \pT indicating strong charm medium interactions at this kinematic region. At the intermediate \pT range ($\sim$0.7-2 GeV/c), the data shows an enhancement which can be described by models including coalescence of charm and light quarks.

We compare our $D^0$ \raa results with those of pions at RHIC (red squares) and D-mesons at LHC (blue triangles)~\cite{star1, star4, star6, alice2}. As shown in Figure 3, at high \pT $R_{AA}(D^0)$ is close to $R_{AA}(\pi)$, which can be explained by taking into account the fact that charm energy loss is an interplay of elastic and radiative energy loss~\cite{theory}. The D-meson \raa at \pT $> 2$ GeV/c is also comparable between RHIC and LHC despite of a factor of 14 difference in collision energy. 
\begin{figure}[htbp]
\hspace{+1.2cm}
\begin{minipage}[t]{0.38\linewidth}
\begin{center}
\includegraphics[width=\textwidth]{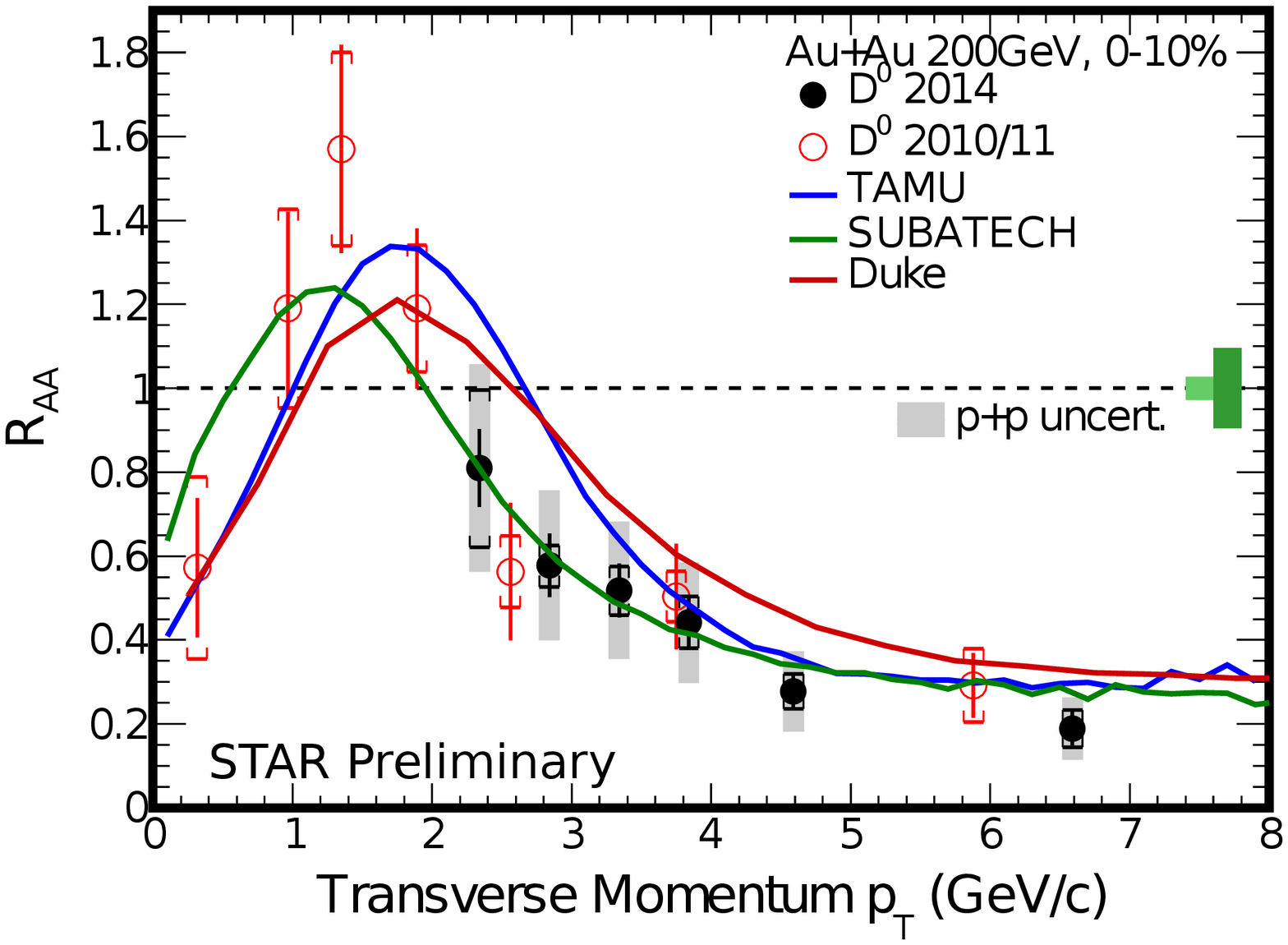}
\end{center}
\end{minipage}
\hspace{+1.2cm}
\begin{minipage}[t]{0.38\linewidth}
\begin{center}
\includegraphics[width=\textwidth]{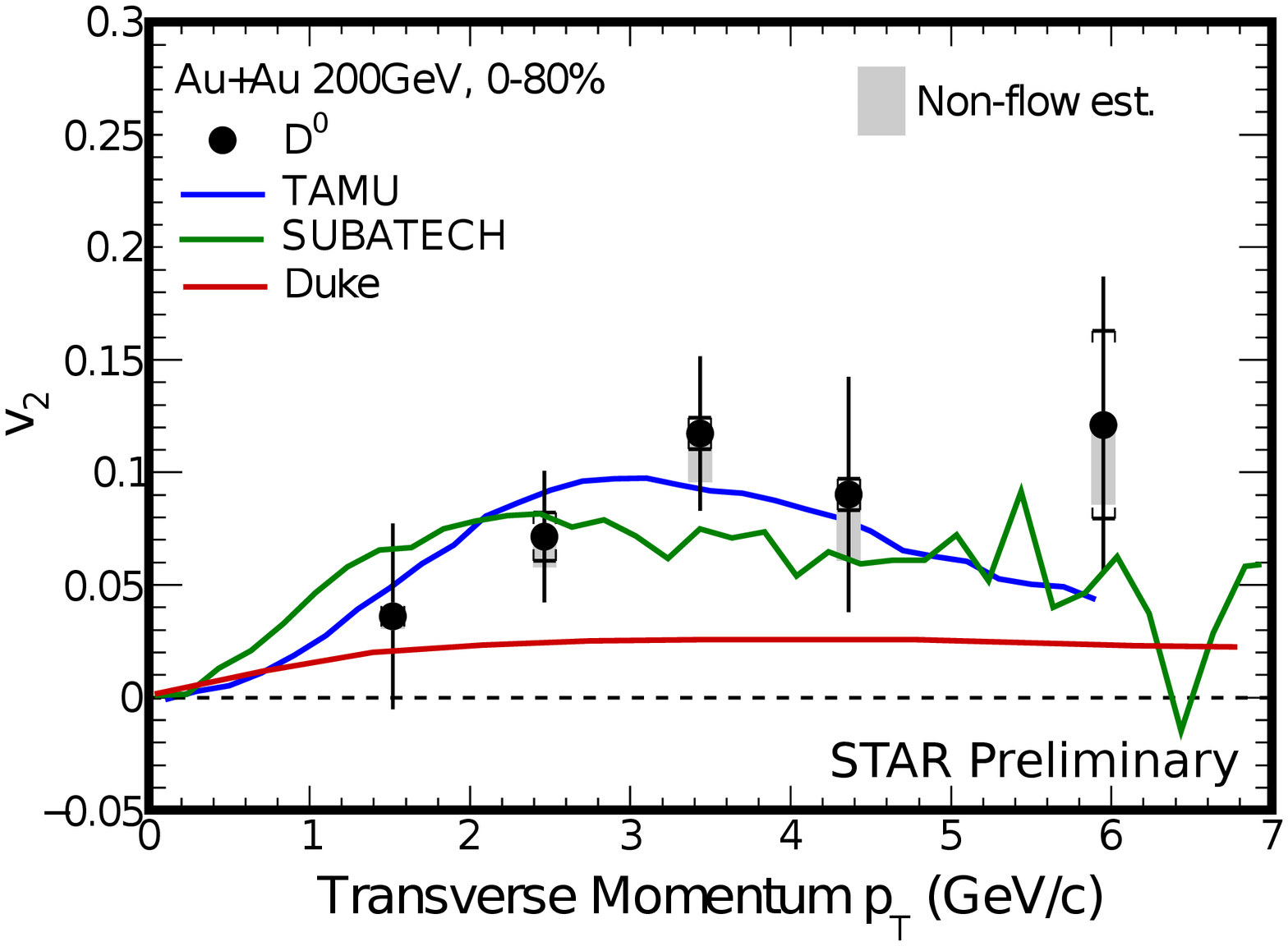}
\end{center}
\end{minipage}
\caption{$D^0$ meson \raa (left) and $v_{2}$ (right) vs. \pT in Au+Au collisions at $\sNN$ = 200 GeV comparing with various model calculations.}
\label{fig:D0Spectramodel}
\end{figure}

The left panel of Fig. 4 shows the \raa results from the most central (0-10\%) \AuAu collisions compared with various model calculations. The Duke model uses a Langevin simulation with an input diffusion coefficient parameter- 2${\pi}TD_{S}= 7$, where $D_S$ is heavy quark spacial diffusion coefficient and T is medium temperature, which was tuned to the LHC D-meson \raa data~\cite{duke, theory}. The TAMU calculation uses a non-perturbative approach and the full T-matrix calculation with internal energy potential, which predicts 2${\pi}TD_{S}$ to be $\sim$ 2-10~\cite{theory}. The SUBATECH group uses the pQCD calculation with Hard Thermal Loop technique which indicates the 2${\pi}TD_{S}$ $\sim$2-4~\cite{theory}. These three models can describe our \raa data points reasonable well. In the meantime, we also compare our first measurement of $D^0$ $v_2$ to model calculations~\cite{star5}. The TAMU and SUBATECH calculations can describe the measured $D^0$ $v_2$ as well, while the specific DUKE calculation with 2${\pi}TD_{S}= 7$ seems to underestimate the $D^0$ $v_2$. Our data favor models with finite charm flow. To further constrain our understanding of the medium diffusion coefficient, it will be beneficial to systematically study each ingredient in different model calculations.

\section{Summary and Outlook}
We report the first measurement of $D^0$ \raa in the most central \AuAu collisions at $\sNN$ = 200 GeV using the STAR newly installed HFT detector. New results confirm the strong suppression at high \pT with a much improved precision. Theoretical models with charm diffusion coefficient 2${\pi}TD_{S}$ $\sim$ 2-12 can reproduce simultaneously both the $D^0$ \raa and $v_2$ data in \AuAu collisions at RHIC. 

In year 2015, STAR has collected high statistics datasets with the HFT in \pp and $p+Au$ collisions at $\sNN$ = 200 GeV. Analyses of these datasets will improve our \pp baseline for the \raa estimation and help understand cold nuclear matter effects. In addition, STAR has requested to collect 2 billion MB \AuAu events in the year 2016 run which will allow more precise determination of the sQGP transport properties. 



%% References

\section*{Acknowledgement}
We express gratitude to RNC group at LBNL and HEPG at USTC for their support. From USTC, the author is supported in part by the NSFC under Grant No. 11375172 and MOST under No. 2014CB845400.

\bibliographystyle{elsarticle-num}
\bibliography{<your-bib-database>}






% \end{thebibliography}

% \section*{References}

\end{document}